\documentclass[prl,floatfix,twocolumn,showpacs,amsmath,amssymb,superscriptaddress,titlepage,natbib]{revtex4-1}
\usepackage{graphicx}
\usepackage{epstopdf}
\usepackage{epsfig}
\usepackage{dcolumn}
\usepackage{bm}
\usepackage{color}
\usepackage{subfigure}
\usepackage{wrapfig}
\usepackage{rotate,color,url}
\usepackage{longtable}
\usepackage{time}
\usepackage[pagebackref=false,colorlinks,linkcolor=blue,citecolor=blue,urlcolor=magenta]{hyperref}

\begin{document}
\newcommand {\be}{\begin{equation}}
\newcommand {\ee}{\end{equation}}
\newcommand {\bea}{\begin{array}}
\newcommand {\cl}{\centerline}
\newcommand {\eea}{\end{array}}
\newcommand {\pa}{\partial}
\newcommand {\al}{\alpha}
\newcommand {\de}{\delta}
\newcommand {\ta}{\tau}
\newcommand {\ga}{\gamma}
\newcommand {\ep}{\epsilon}
\newcommand {\si}{\sigma}
\newcommand{\up}{\uparrow}
\newcommand{\down}{\downarrow}

\title{Complexity of eye fixation duration time series in reading of Persian texts: A multifractal detrended fluctuation analysis}

\author{Mohammad Sharifi}
\affiliation{Department of Physics, Isfahan University of Technology, Isfahan 84156-83111, Iran}

\author{Hamed Farahani}
\affiliation{Department of Physics, Isfahan University of Technology, Isfahan 84156, Iran}

\author{Farhad Shahbazi}
\affiliation{Department of Physics, Isfahan University of Technology, Isfahan 84156, Iran}
\affiliation{School of Physics, Institute for Research in Fundamental Sciences (IPM), Tehran 19395, Iran}

\author{Masood Sharifi}
\affiliation{Department of Psychology, Shahid Beheshti University, Tehran 19839, Iran}

\author{Christofer T. Kello}
\affiliation{School of Social Sciences, Humanities and Arts, University of California, Merced, 95343, USA}

\author{Marzieh Zare}
\affiliation{School of Computer Science, Institute for Research in Fundamental Science, Tehran 19395, Iran}

\date{\today}

\begin{abstract}

There is growing evidence that cognitive processes may have fractal structures as a signature of complexity. It is an an ongoing topic of research to study the class of complexity and how it may differ as a function of cognitive variables. Here, we explore the eye movement trajectories generated during reading different Persian texts. Features of eye movement trajectories were recorded during reading Persian texts using an eye tracker. We show that fixation durations, as the main components of eye movements reflecting cognitive processing, exhibits multifractal behavior. This indicates that multiple exponents are needed to capture the neural and cognitive processes involved in decoding symbols to derive meaning. We test whether multifractal behavior varies as a function of two different fonts, familiarity of the text for readers, and reading silently or aloud, and goal-oriented versus non-goal-oriented reading. We find that, while mean fixation duration is affected by some of these factors, the multifractal pattern in time series of eye fixation durations did not change significantly. Our results suggest that multifractal dynamics may be intrinsic to the reading process. 

\end{abstract}

\maketitle

\section{Introduction}

Reading has been broadly studied across different alphabetic and logographic systems, such as English, Chinese, Arabic, Japanese, French or German in varieties of techniques and behavioral tests~\cite{reynolds1995robust, rayner1998eye}. The perceptual process of reading is reflected in reaction times or eye movements, and a cognitive outcome is text comperhension. Eye movements during reading are generated by complex self-regulating systems that process inputs from different regions of the brain. Eye movements are extremely heterogeneous and non-stationary, two properties that may arise from complex underlying dynamics of the task. Furthermore, as reading is a cognitive process, complexity arising from the cognitive load depends on the context, textual, and typographical variables.

Reading involves intrinsic and extrinsic factors. Extrinsic factors include textual and typographical variables such as font complexity and the difficulty of the text, while intrinsic factors are general across stimulus and the context originating from processes and structures that change only on longer timescales of reading skills, learning, and development. 

The complexity of reading may be expressed in terms of fractal patterns generated by eye movement characteristics such as fixation duration. Cognitive functions like visual search \cite{foraging}, scene perception~\cite{zelinsky2005role}و and visual foraging \cite{rhodes2014intrinsic} have revealed evidence for heavy-tailed distributions~\cite{kello2007emergent, rhodes2014intrinsic}. Spatial clustering of eye movements follow a power law distribution, saccade length distributions are log-normally distributed, and the speeds of slow, small amplitude movements occurring during fixations follow a $1/f$, spectral power law relation~\cite{kello2008pervasiveness, kello2013critical}, suggesting a self-similar pattern for eye fixation duration dynamics in a reading task.

Previous studies of heavy tails in reading have focused on monofractal analyses; assuming that distributions can be captured by a single scaling exponent. To test this assumption, a multifractal analysis is required for the fixation duration time series during different reading experiments. Indeed, a fractal, which is self-similar in all the scales, is simply described by a power law with the same exponent at all scales. However, to characterize a more complex pattern, i.e. a multifractal, a continuous range of exponents is required~\cite{sornette2006critical, feder2013fractals}, 

So far, few studies has focused on the fractal and multifractal patterns of eye movement during reading.  Van Orden et. al, \cite{vanorden}, referring to the folk expression that "the eyes are the windows to the soul" employed fractal and multifractal methods to find whether eye-movments detected by an eye-tracker generates intrinsic random variation and how features of the data recording procedure affected the structure measurement variability. Their results revealed that the structure of variation from a fake eye was random and uncorrelated in contrast to the fractal structure from a fixated, real human eye. Furthermore, it was demonstrated that data-averaging generally changes the structure of variation, introducing spurious structure into eye movement variability. 

The aim of the present study is to investigate the roles of extrinsic and intrinsic factors on the complexity of eye movements during reading the Persian texts. In four experiments, the participants read texts with different content and font complexity. The fixation duration time series of the eye movements were recorded by an eye-tracker and both standard statistics as well as multifractal detrended fluctuation analyses (MF-DFA) are used to investigate them.

\section{materials and methods}
\label{experiments}

\subsection{Ethics}
The study was approved by the institutional review board of the University of Shahid Beheshti. Before the beginning of the study, participants were presented with a written consent form and written consent was obtained from each individual.

\subsection{Apparatus}
Each participant was seated approximately 20 inches in front of a 22-inch flat panel LCD monitor. Participants viewed each of the 4 images in random order for 45 seconds per image. Monocular eye positions were recorded at 120 Hz using an Eye Link II head mounted eye-tracker. Participants were instructed to view each image for 45 s. The device also calculates and reports the eye fixation errors, therefore the trials that are high in error can be detected and removed from the analysis.

\subsection{Stimuli}
\noindent \textbf{{\em Familiarity}}. The familiar text was from the Psychology literature that is generally known to Psychology students. The unfamiliar text was taken from Physics literature whose context is not known for an average number of psychology students. The familiar and unfamiliar texts were selected based on some features such as the short word length (3 to 5 characters) and the long word length (6 characters and more). Therefore, both familiar and unfamiliar texts consisted of short and long words in an equal proportion.

\noindent \textbf{{\em Font}}. The texts were written in two font types: first the Lotus font, commonly used for books using, and the other was the Pen font that is similar to the handwritten font. Texts were typed using @Microsoft Word. 
Selected texts matched in terms of difficulty, context and word frequency. Overall, four different stimuli were used in the current study: a) familiar text written in Lotus font, b) familiar text written in Pen font, c) unfamiliar text written in Lotus font, d) unfamiliar text written in Pen font. 

\subsection{Procedure}
At the beginning of the study, two texts were prepared according to stimuli section. Then a pilot study was conducted on 10 participants to examine the text properties, in terms of familiarity and unfamiliarity level, coherence between the two texts in Lotus and Pen fonts, and performance of research instruments. This pilot study led to some amendments to the size and content of the texts.
Also, in the calibration stage, the distance between subject's eyes and the eye-tracking device was adjusted. The degree of accuracy and reliability was performed using the eye-tracker device, and in case the error rate was low, the participants were tested by the device in one session and in four successive phases.
Each participant was seated in a quiet experimental room about 20 inches in front of a computer monitor placed on a table. For each trial, a text written either in Lotus or Pen font appeared on the monitor. Participants were instructed to read the text. Trials included Silent Not Goal-Oriented (SNGO), Silent Goal-Oriented (SGO), and Loud Not Goal-Oriented (LNGO).
The familiar text in the Lotus and Pen fonts were read by the subject in the first and the second phase respectively, and the unfamiliar text in the Lotus and Pen fonts were read by them in the third and the fourth phases, respectively. Throughout all phases, the subject's eye movements were recorded by an eye-tracker device while reading the texts. The device records the fixation durations, saccades, regressions and overall time spent to read every word of the text. Prior to the test, subjects were asked to follow the test procedure in accordance with written instructions, and read the text at their own pace. They were also instructed to find the next texts shown on the screen by pressing the spacebar on the keyboard in front of them.

\subsection {Experiment I: Silent Not Goal-Oriented (SNGO)}
{\em Participants}. 16 undergraduate students were recruited from Shahid Beheshti University, Psychology Department. They all had normal or corrected vision. Since the eye-tracker device in the laboratory was fixed and impossible to move, the familiar text contained psychological knowledge, the psychology undergraduate students who were often easy to reach were selected as participants. They were screened for potential issues that might affect sampling rates, such as eyeglasses or cosmetics. The participants were asked to read both familiar and unfamiliar texts silently. The experiment was not goal oriented, so they were not supposed to answer any questions relating to the texts. 

\subsection{Experiment II: Silent Goal-Oriented (SGO) }

{\em Participants}. 16 participants were recruited from Shahid Beheshti University, Psychology Department. They were all right-handed, and had normal or corrected vision. Participants were screened for potential issues that might affect sampling rates, such as eyeglasses or cosmetics. Since the procedure was goal-oriented, a researcher-made test consisting of 8 multiple choice questions. 2 questions per text, with the same difficulty level, was extracted from the four stimuli and was distributed among the participants after the experiment. Scores were not taken into account in any of the analyses. 

\subsection{Experiment III. Loud Not Goal-Oriented (LNGO)}

{\em Participants}. 16 participants were recruited from Shahid Beheshti University, Psychology Department. They were all right-handed, and had normal or corrected vision. Participants were screened for potential issues that might affect sampling rate, such as eyeglasses or cosmetics. The participants were asked to read both familiar and unfamiliar texts loudly. The experiment was not goal-oriented, so they were not supposed to answer any questions relating to the texts. 


\begin{figure*}
\includegraphics[width=\textwidth]{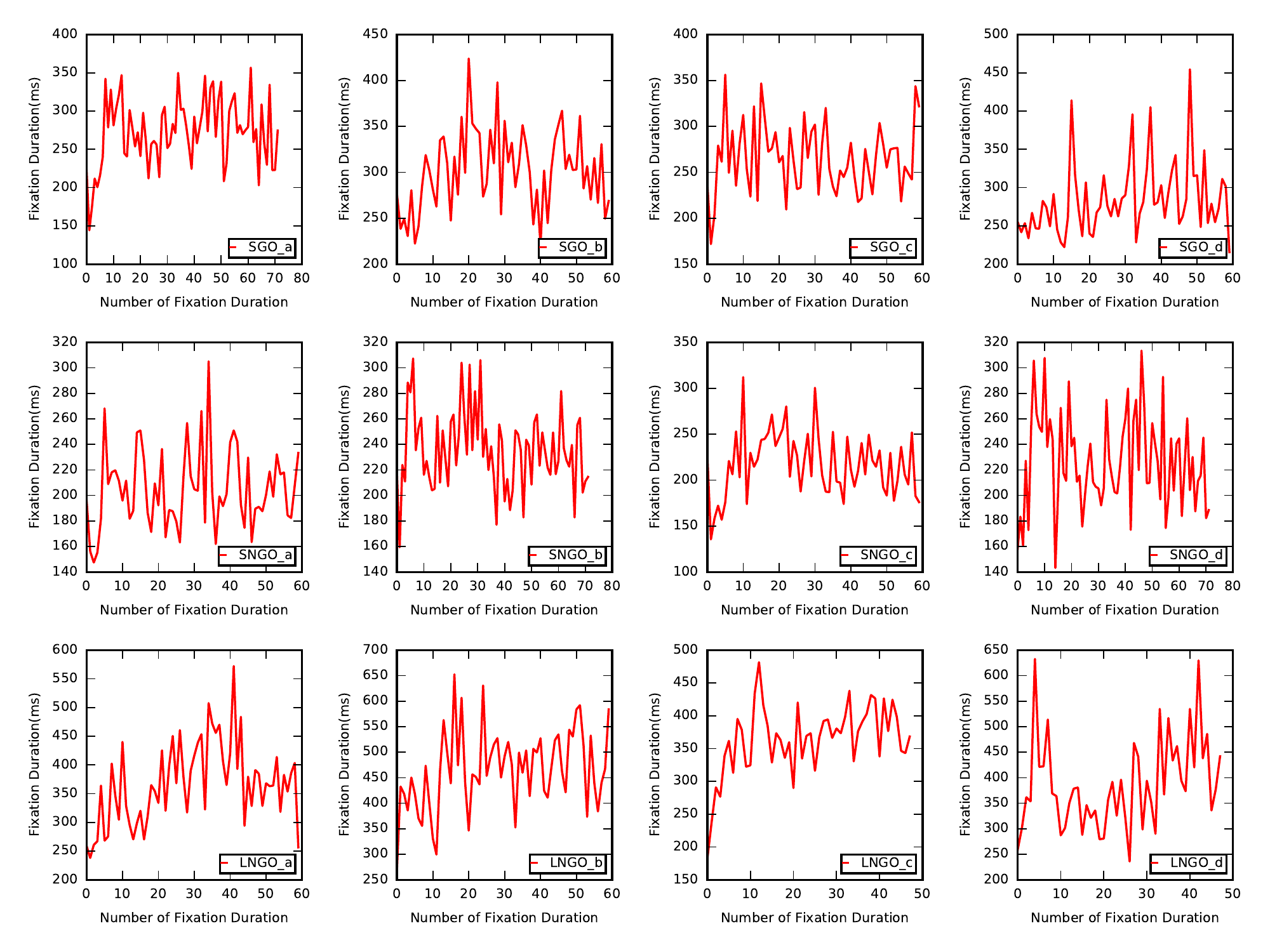}

\caption{(Color online) Fixation duration time series obtained in the eye-track record of one subject in each experiment. ({\bf Left}) Silent goal oriented (SGO) with (a) familiar lotus, (b) familiar Pen, (c) unfamiliar Lotus and (d) unfamiliar Pen text. ({\bf Middle}) Silent non goal oriented (SNGO) with (a) familiar lotus, (b) familiar Pen, (c) unfamiliar Lotus and (d) unfamiliar Pen text. ({\bf Right}) Load non-goal oriented (LNGO) with (a) familiar lotus, (b) familiar Pen, (c) unfamiliar Lotus and (d) unfamiliar Pen text.}
\label{series}
\end{figure*}

\section{Method: multifractal detrended fluctuation analysis (MF-DFA)}
\label{method}

To determine the complexity of fixation duration time series, we use MF-DFA. This method, initially proposed by Kantelhardt et al.~\cite{kantelhardt2002multifractal,kantelhardt2012fractal}, finds many applications in time series such as sunspot variations~\cite{movahed2006multifractal}, traffic data~\cite{shang2008detecting}, economical data~\cite{lim2007multifractal,yuan2009measuring,wang2009analysis}, heart rate time series~\cite{galaska2008comparison}, human brain electrical signals~\cite{zorick2013multifractal}, streamflow~\cite{zhang2008multifractal} and wind~\cite{telesca2011analysis} records. 

The procedure of MF-DFA is as follows. Consider the time series $x(i)$ of length $N$, its cumulative series is defined as

\begin{equation}
Y(i)=\sum_{k=1}^{i} (x_k-\langle x\rangle); ~~ i=1,2,\cdots N,
\end{equation}

in which $\langle x\rangle$ is the average of the time series over the whole range of measured data. Indeed, $Y(i)$ can be assumed as a random walk whose steps are the fluctuations of the time series around its average at each time. In the next step, we proceed to remove the local trends from the profile series $Y$. For this, we divide, $Y(i)$ series to the segments of length $dl$ and carry a least-square fit on the data in each segment by a polynomial $Y_{\nu}$. The variance of data in each segment around the fitted polynomial is given by 

\begin{equation}
F^{2}(dl,\nu)={1\over s}\sum_{i=1}^{s}\{ Y[(\nu-1)dl+i]-Y_{\nu}(i)\}^{2}; ~~ \nu=1,2,\cdots N_s
\end{equation}
where $N_s=N/dl$. In this work, we choose the linear polynomials for detrending the data, i.e. first order DFA or DFA1. 

After detrending, we move on to investigate the multifractal properties of the time series. To this end, we define the $q$-th fluctuation moment for any real $q\neq 0$ as 
\begin{equation}\label{fq}
F_{q}(dl)=\left( \frac{1}{N_s} \sum_{\nu=1}^{N_s}[F^{2}(dl,\nu)]^{q/2}\right)^{1/q}.
\end{equation}
For $q=0$, $F_{0}(dl)$ can be written as 
\begin{equation}
F_{0}(dl)=\exp \left( \frac{1}{2N_s} \sum_{\nu=1}^{N_s}\ln (F^{2}(dl,\nu))\right).
\end{equation}
It can be shown that for large enough $dl$, $F_{q}(dl)$ obeys the following scaling law

\begin{equation}
F_{q}(dl)\sim s^{h(q)},
\end{equation}

where $h(q)$ is called {\em generalized Hurst exponent}. For a monofractal time series fluctuations are homogenous over all the segments at each scale, hence the statistics of fluctuations is similar in small and large scales making $h(q)$ an independent function of $q$. For $q=2$, $h(2)$ is the same as standard Hurst exponent, $H$. For uncorrelated time series, we have $H=1/2$, $H < 1/2$ for anti-correlated, and $H>1/2$ for correlated time series~\cite{feder2013fractals}. 
For a multifractal series, however, the fluctuations do not behave similarly in all the segments. For positive value of $q$ the segments with stronger fluctuations plays a major rule in the summation of Eq.\eqref{fq}. This makes the scaling of fluctuations at small scales be different from the one at large scales, and as a result, making $h(q)$ a $q$-dependent function in this case. 

To obtain a singularity spectrum from MF-DFA-extracted from standard MF analysis~\cite{feder2013fractals}, one needs to calculate the local singularity strength or the H{\"o}lder exponent $\alpha$ from the generalized Hurst exponent $h(q)$. It can be shown that the local singularity strength $\alpha$ corresponding to the power $q$, can be obtain by 

\begin{equation}\label{tauq}
\alpha(q)=\frac{d\tau(q)}{dq},
\end{equation}
in which $\tau(q)$ is the mass exponent and is related to $h(q)$ as
\begin{equation}
\tau(q)=qh(q)-1.
\end{equation}

\noindent Eq.\eqref{tauq} implies that for each value of $q$, there is a unique exponent $\alpha(q)$ which indicates the singularity exponent of the segments in the time series which dominantly contribute in $F_q(s)$. Indeed, the exponent $\alpha$ quantifies the speed of the vanishing fluctuations versus the decreasing length of the segments in which the measure is defined. Dependence of the measure on the segment length is given by a power law as $p_{n}\sim s^{\alpha_n}$, where $p_{n}$, $\alpha_n$ and $s$ are the local fluctuation measure, local H{\"o}lder exponent, and the corresponding segment length, respectively. 
$\alpha_n$ is positive, therefore the larger the value of $\alpha_n$, the faster the decay of fluctuations is as $s\rightarrow 0$; in other words, the time series in the corresponding region is smoother. Therefore, the large(small) value of $\alpha$ indicates the smooth(rough) segments in the time series. 

Each $q$ exponent magnifies only the segments with the singularity strength $\alpha(q)$ in Eq.\eqref{fq}, and 
the segments in the time series whose singularity is characterized by 
$\alpha(q)$, form a fractal subset with a fractal dimension defined by $f(\alpha)$ is given by 
\begin{equation}
f(\alpha)=q\alpha(q)-\tau(q(\alpha)),
\end{equation}
which is the Legendre transformation of $\tau(q)$~\cite{feder2013fractals, sornette2006critical}. It can be shown that $f(\alpha)$ is a convex up function of $\alpha$ with a maximum equal to $1$ for any $1D$ time series. Indeed, the maximum of $f(\alpha)$ which corresponds to $q=0$, indicates the capacity or topological dimension of the time series which is always equal to $1$ for $1D$ time series. The width of singularity spectrum $f(\alpha)$ which is defined as $\Delta\alpha=\alpha_{\rm max}-\alpha_{\rm min}$, shows the range of the H{\"o}lder exponents required to describe the time series, hence, the wider the range of $\alpha$, the more complex the time series will be. 

\section{results and discussion}
\label{results}

\begin{table}
\caption{The average mean fixation time corresponding to the different fixation time series. The numbers in the parentheses indicate the uncertainty of the mean value calculated by the standard deviation over different subjects in each experiment.}
\label{tab:mean}
\begin{ruledtabular}
\begin{tabular}{|c|c|c|c|}
& SGO (ms) & SNGO (ms) & LNGO (ms) \\ \hline 
(a)-Familiar-Lotus & 266(23) & 198(22) & 355(29) \\ \hline
(b)-Familiar-Pen & 294(30)& 226(17) & 468(37) \\ \hline
(c)-Unfamiliar-Lotus & 260(19) & 211(16) & 361(28) \\ \hline
(d)-Unfamiliar-Pen & 271(23) & 202(15) & 399(36) 
\end{tabular}
\end{ruledtabular}
\end{table}

\begin{figure}[t]
\includegraphics[width=\columnwidth]{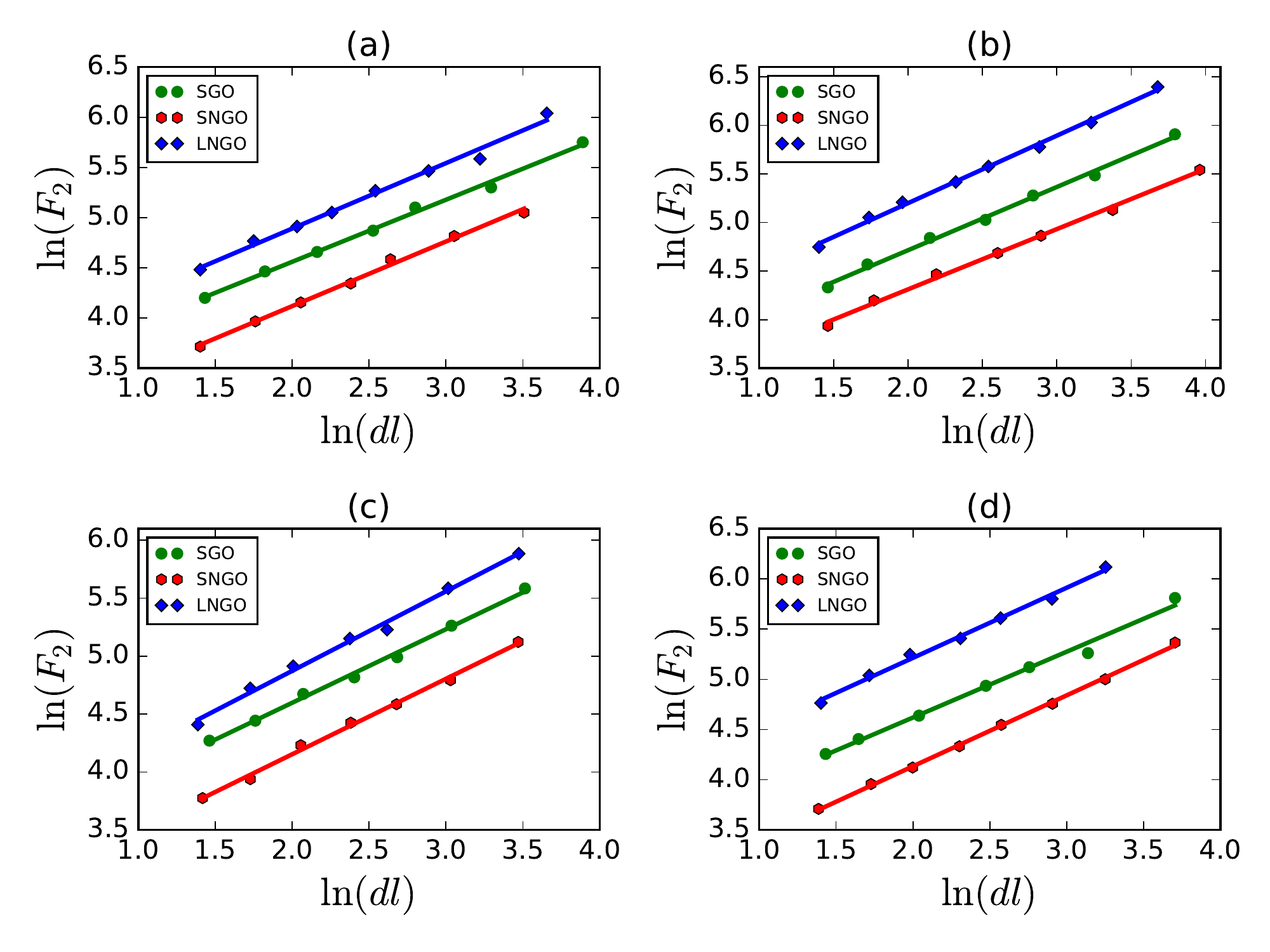}
\caption{(Color online) $\ln(F_2)$ versus $\ln(dl)$ averaged over the different reading tasks SGO, SNGO and LNGO on the four texts, ({\bf top left}) familiar with Lotus font, ({\bf top right}) familiar with Pen font, ({\bf bottom left}) unfamiliar with Lotus font and ({\bf bottom right}) unfamiliar with Pen font. The slope of each line gives the Hurst exponent of the corresponding experiment.}
\label{Hurst}
\end{figure}

\begin{table}
\caption{The average Hurst exponents corresponding to the different fixation time series. The numbers in the parentheses indicate the uncertainty of the mean Hurst exponents.}
\label{tab:H}
\begin{ruledtabular}
\begin{tabular}{|c|c|c|c|}
& SGO & SNGO & LNGO \\ \hline 
(a)-Familiar-Lotus & 0.62(0.03) & 0.65(0.02) & 0.65(0.04) \\ \hline
(b)-Familiar-Pen & 0.65(0.03)& 0.62(0.02) & 0.70(0.02) \\ \hline
(c)-Unfamiliar-Lotus & 0.64(0.02) & 0.66(0.02) & 0.69(0.03) \\ \hline
(d)-Unfamiliar-Pen & 0.66(0.04) & 0.70(0.01) & 0.70(0.03) 
\end{tabular}
\end{ruledtabular}
\end{table}

\begin{figure}
\includegraphics[width=\columnwidth]{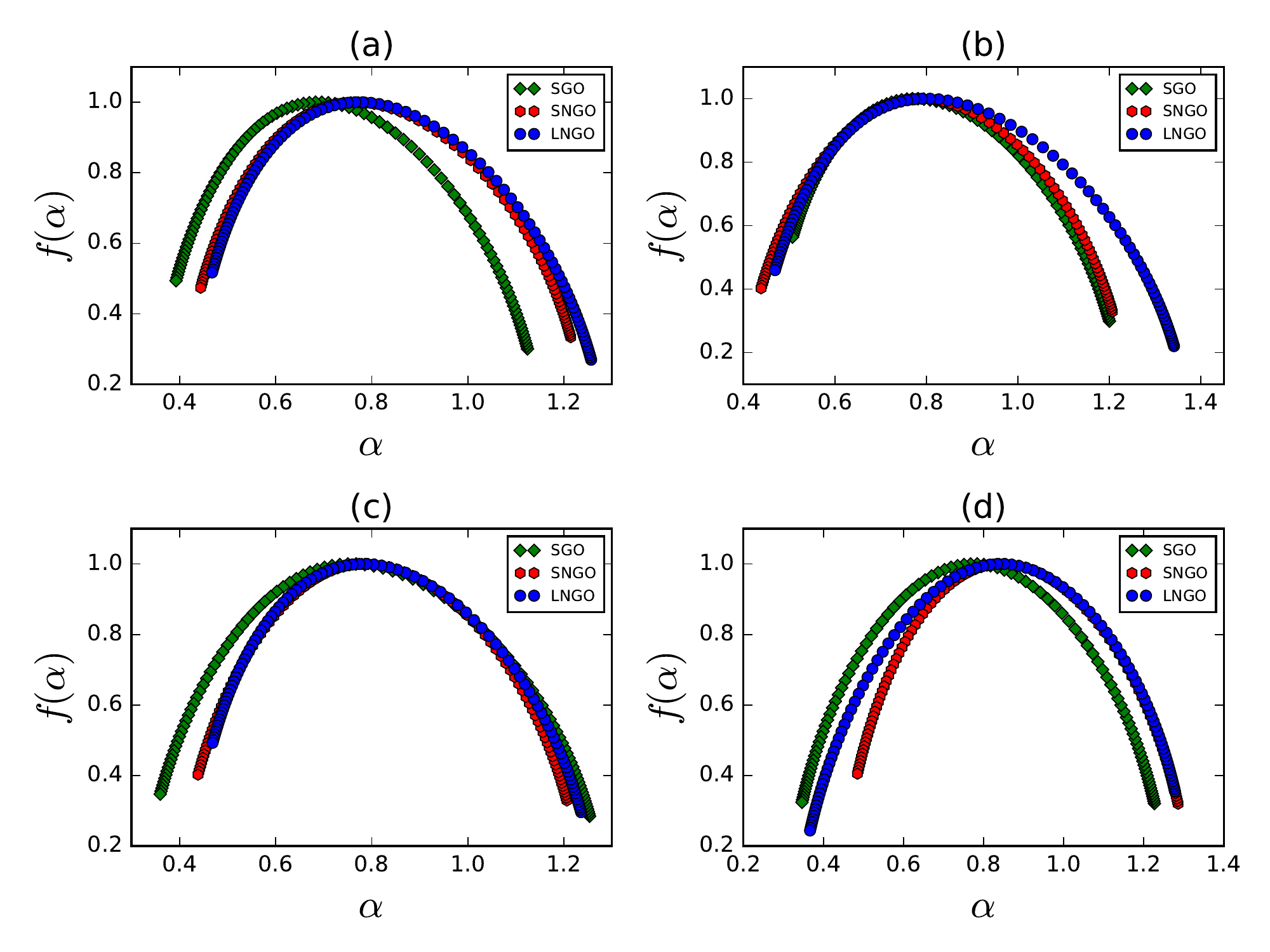}
\caption{(Color online) The singularity spectrums corresponding to the different reading tasks SGO, SNGO and LNGO on the four texts, ({\bf top left}) familiar with Lotus font, ({\bf top right}) familiar with Pen font, ({\bf bottom left}) unfamiliar with Lotus font and ({\bf bottom right}) unfamiliar with Pen font.}
\label{mfdfa1}
\end{figure}

\begin{figure}
\includegraphics[width=\columnwidth]{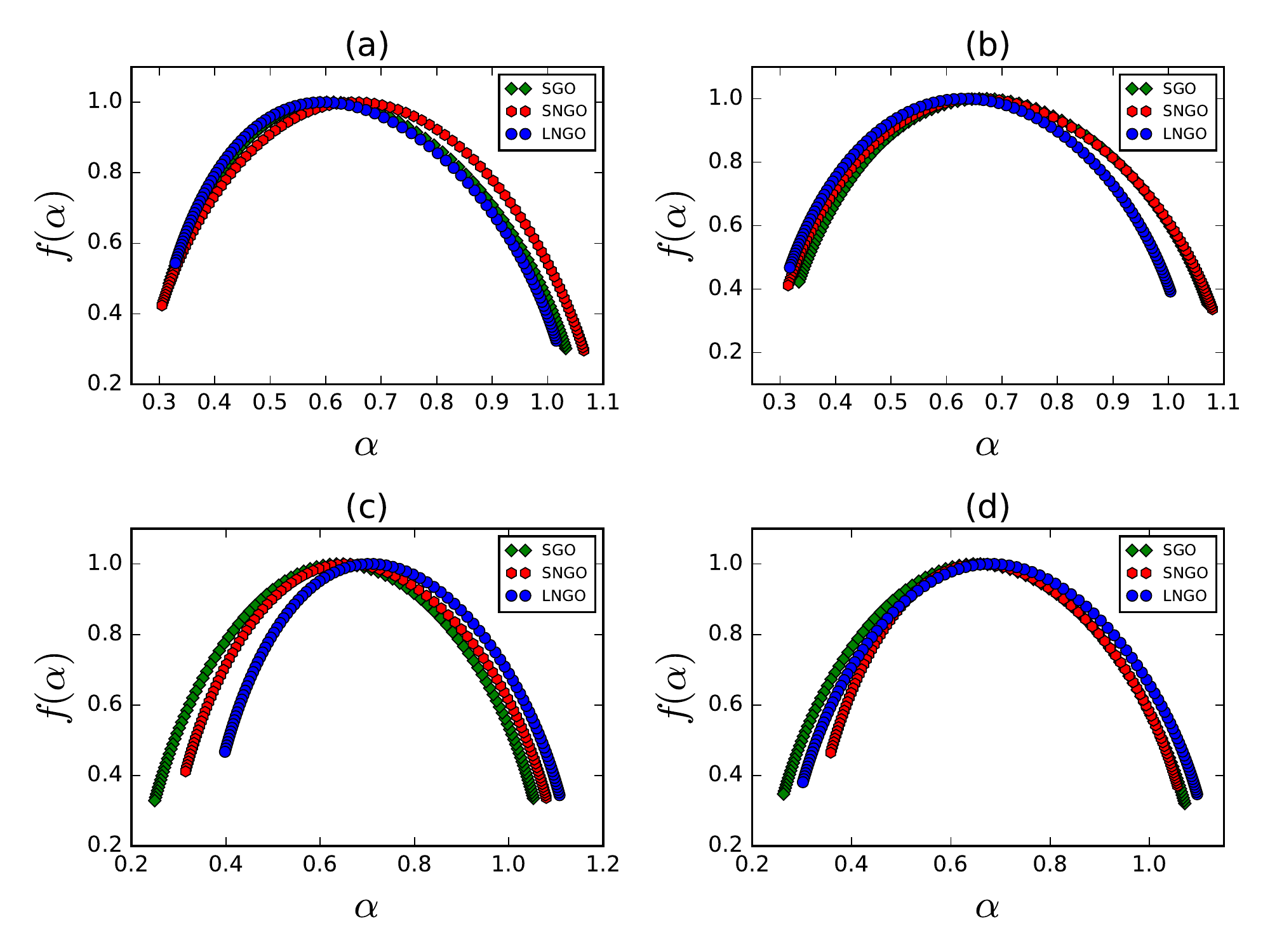}
\caption{(Color online) The singularity spectrums of the shuffled fixation time series corresponding to different reading tasks SGO, SNGO and LNGO on the four texts, ({\bf top left}) familiar with Lotus font, ({\bf top right}) familiar with Pen font, ({\bf bottom left}) unfamiliar with Lotus font and ({\bf bottom right}) unfamiliar with Pen font.}
\label{mfdfas}
\end{figure}

\begin{table}
\caption{The average width of singularity spectrum ($\Delta\alpha$) corresponding to the different fixation time series. The numbers in the parentheses indicate the uncertainty of the mean widthes. }
\label{tab:alpha}
\begin{ruledtabular}
\begin{tabular}{|c|c|c|c|}
& SGO & SNGO & LNGO \\ \hline 
(a)-Familiar-Lotus & 0.73(0.04) & 0.77(0.04) & 0.78(0.04) \\ \hline
(b)-Familiar-Pen & 0.69(0.03)& 0.76(0.04) & 0.9(0.05) \\ \hline
(c)-Unfamiliar-Lotus & 0.90(0.05) & 0.76(0.04) & 0.76(0.04) \\ \hline
(d)-Unfamiliar-Pen & 0.90(0.05) & 0.80(0.04) & 0.9(0.05) 
\end{tabular}
\end{ruledtabular}
\end{table}
\begin{table}
\caption{The average kurtosis for the different fixation duration time series. The numbers in the parentheses indicate the uncertainty of the mean kurtosis values.}
\label{tab:k}
\begin{ruledtabular}
\begin{tabular}{|c|c|c|c|}
& SGO & SNGO & LNGO \\ \hline 
(a)-Familiar-Lotus & 2.3(1.4) & 2.1(1.2) & 1.3(0.7) \\ \hline
(b)-Familiar-Pen & 2.7(1.8)& 2.7(1.1) & 2.2(1.1) \\ \hline
(c)-Unfamiliar-Lotus & 5.3(2.5) & 3.7(1.7) & 2.0(1.1) \\ \hline
(d)-Unfamiliar-Pen & 6.0(3.5) & 3.3(1.4) & 6.6(2.8) 
\end{tabular}
\end{ruledtabular}
\end{table}

The fixation duration time series (the time period in which eyes remains fixed on a part of text) of randomly selected subjects are depicted in Fig.~\ref{series} for different experiments. The mean values of fixation durations in each experiment, together with their statistical errors, are given in Table~\ref{tab:mean}. This table shows that in all the reading experiments SGO, SNGO and LNGO, there were no meaningful differences between average mean fixation times for different stimuli. However, the mean fixation time was the largest for LNGO and the smallest for SNGO, which indicates that reading aloud prolonged fixation times compared to silent reading. 

To gain some insight into the correlation of the fixation duration time series, we calculated the Hurst exponent. The Hurst exponents averaged over all the subjects in each reading experiment are listed in Table~\ref{tab:H}. It can be seen that the Hurst exponents for the fixation time series varies from $\sim 0.6$ to $\sim 0.7$, which is an indication of long-range correlations in eye fixation durations for all experiments. However, there were no reliable differences in Hurst exponents across experiments. 

Next, we proceed to investigate the complexity of the time series by calculating the singularity spectrum using MF-DFA method.
Fig.~\ref{mfdfa1} illustrates the singularity spectra $f(\alpha)$ obtained for the three reading experiments SGO, SNGO and LNGO on the four stimuli including: (a)-familiar with Lotus font, (b)-familiar with Pen font, (c)-unfamiliar with Lotus font and (d)-unfamiliar with Pen font. The complexity of each time series is encoded in the width of singularity spectrum $f(\alpha)$. 

Time series with wider spectra contain more singularity strengths $\alpha$, meaning that they contains more interwoven fractal subsets, and hence more complexity. 
Within the statistical errors, the average widths of singularity spectra, $\Delta\alpha$, given in Table~\ref{tab:alpha}, is $\sim0.8$ for all the experiments.  Thus we found no reliable differences in terms of the complexity of the fixation duration time series across the three different reading experiments. 

As a final analysis, to examine the effect of correlation on the multifractality of fixation duration time series, we shuffle the time series to eliminate the correlations, and calculate the singularity spectrum. The results are illustrated in Fig.\ref{mfdfas}, and show that shuffling only tends to shift the whole spectra toward the lower values of $\alpha$.  However, the width of $\Delta\alpha$ remains almost unchanged. This analysis shows that the correlations have minor effect on the multifractality, which suggests that multifractality arose from large fluctuations in fixation durations. To test this hypothesis, we calculate the kurtosis for each time series

\begin{equation}
k=\frac{\langle x^4 \rangle}{\langle x^2 \rangle^2}-3.
\end{equation} 
$X_i$ is the time series. The kurtosis vanishes to a Gaussian distribution and its deviation from zero is an evidence of the presence of fluctuations larger than the Gaussian limit. The average values of kurtosis for the fixation duration time series are summarized in Table~\ref{tab:k}, showing relatively remarkable deviations from normality. This analysis leaves open the possibility that multifractal patterns appeared because of finite lengths of the recorded data. Nevertheless, the large kurtosis obtained suggests that multifractality was at least partly due to high volatilities in the intrinsic dynamics of the eye movement. 

\section{conclusion}
In summary, we performed statistical analyses on the fixation duration time series obtained by tracking the eye movements in three reading experiments and for four different stimuli. We found that reading aloud, even without any goal increases the mean fixation duration compared to reading silently. Therefore, assuming that fixation duration is proportional to the cognitive load in the brain, suggesting that simultaneous activation of the brain regions responsible for speech and visual processing slows down the reading speed.

Using DFA method, we calculated the Hurst exponents as well as the multifractal singularity spectrum of the data. We found that in all the experiments, the fixation duration time series are positively correlated and display a multifractal pattern. Removing the correlations by shuffling the data, however did not significantly affect the singularity spectra. Moreover, fluctuations beyond the Gaussian limit indicates that the multifractality would be intrinsic; revealing the high volatility in the eye track dynamics. Surprisingly, while several studies have emphasized on the importance of properties of the stimulus on eye movemenet behaviors \cite{Findlay, Itti,Lamy}, we did not find any reliable difference in the multifractality of data in the different reading experiments, supporting \cite{Land,Underwood,Ballard}. The fact that the singularity spectra does not vary in different experiments, leads us to the conclusion that the eye movement dynamics is developed in such a way that enables eye to easily response to wide range of external stimuli. This results also support the findings of \cite{PITS:PITS20543}, in which no significant difference was found between aloud and silent reading comprehension, which suggests that reading comprehension can be measured accurately under either reading condition.

\section{acknowledgment}
\noindent M.SH., H.F., M.Z, and F.SH gratefully acknowledge financial support from Iranian Cognitive Science and Technologies council through Grant No.2690.

\bibliographystyle{apsrev4-1}

\bibliography{biblioeyetrack}

\begin{thebibliography}{29}%
\makeatletter
\providecommand \@ifxundefined [1]{%
 \@ifx{#1\undefined}
}%
\providecommand \@ifnum [1]{%
 \ifnum #1\expandafter \@firstoftwo
 \else \expandafter \@secondoftwo
 \fi
}%
\providecommand \@ifx [1]{%
 \ifx #1\expandafter \@firstoftwo
 \else \expandafter \@secondoftwo
 \fi
}%
\providecommand \natexlab [1]{#1}%
\providecommand \enquote  [1]{``#1''}%
\providecommand \bibnamefont  [1]{#1}%
\providecommand \bibfnamefont [1]{#1}%
\providecommand \citenamefont [1]{#1}%
\providecommand \href@noop [0]{\@secondoftwo}%
\providecommand \href [0]{\begingroup \@sanitize@url \@href}%
\providecommand \@href[1]{\@@startlink{#1}\@@href}%
\providecommand \@@href[1]{\endgroup#1\@@endlink}%
\providecommand \@sanitize@url [0]{\catcode `\\12\catcode `\$12\catcode
  `\&12\catcode `\#12\catcode `\^12\catcode `\_12\catcode `\%12\relax}%
\providecommand \@@startlink[1]{}%
\providecommand \@@endlink[0]{}%
\providecommand \url  [0]{\begingroup\@sanitize@url \@url }%
\providecommand \@url [1]{\endgroup\@href {#1}{\urlprefix }}%
\providecommand \urlprefix  [0]{URL }%
\providecommand \Eprint [0]{\href }%
\providecommand \doibase [0]{http://dx.doi.org/}%
\providecommand \selectlanguage [0]{\@gobble}%
\providecommand \bibinfo  [0]{\@secondoftwo}%
\providecommand \bibfield  [0]{\@secondoftwo}%
\providecommand \translation [1]{[#1]}%
\providecommand \BibitemOpen [0]{}%
\providecommand \bibitemStop [0]{}%
\providecommand \bibitemNoStop [0]{.\EOS\space}%
\providecommand \EOS [0]{\spacefactor3000\relax}%
\providecommand \BibitemShut  [1]{\csname bibitem#1\endcsname}%
\let\auto@bib@innerbib\@empty
\bibitem [{\citenamefont {Reynolds}\ and\ \citenamefont
  {Rose}(1995)}]{reynolds1995robust}%
  \BibitemOpen
  \bibfield  {author} {\bibinfo {author} {\bibfnamefont {D.~A.}\ \bibnamefont
  {Reynolds}}\ and\ \bibinfo {author} {\bibfnamefont {R.~C.}\ \bibnamefont
  {Rose}},\ }\href@noop {} {\bibfield  {journal} {\bibinfo  {journal} {IEEE
  transactions on speech and audio processing}\ }\textbf {\bibinfo {volume}
  {3}},\ \bibinfo {pages} {72} (\bibinfo {year} {1995})}\BibitemShut {NoStop}%
\bibitem [{\citenamefont {Rayner}(1998)}]{rayner1998eye}%
  \BibitemOpen
  \bibfield  {author} {\bibinfo {author} {\bibfnamefont {K.}~\bibnamefont
  {Rayner}},\ }\href@noop {} {\bibfield  {journal} {\bibinfo  {journal}
  {Psychological bulletin}\ }\textbf {\bibinfo {volume} {124}},\ \bibinfo
  {pages} {372} (\bibinfo {year} {1998})}\BibitemShut {NoStop}%
\bibitem [{\citenamefont {Wilming}\ \emph {et~al.}(2013)\citenamefont
  {Wilming}, \citenamefont {Harst}, \citenamefont {Schmidt},\ and\
  \citenamefont {König}}]{foraging}%
  \BibitemOpen
  \bibfield  {author} {\bibinfo {author} {\bibfnamefont {N.}~\bibnamefont
  {Wilming}}, \bibinfo {author} {\bibfnamefont {S.}~\bibnamefont {Harst}},
  \bibinfo {author} {\bibfnamefont {N.}~\bibnamefont {Schmidt}}, \ and\
  \bibinfo {author} {\bibfnamefont {P.}~\bibnamefont {König}},\ }\href@noop {}
  {\bibfield  {journal} {\bibinfo  {journal} {PLOS Computational Biology}\
  }\textbf {\bibinfo {volume} {9}},\ \bibinfo {pages} {1} (\bibinfo {year}
  {2013})}\BibitemShut {NoStop}%
\bibitem [{\citenamefont {Zelinsky}\ \emph {et~al.}(2005)\citenamefont
  {Zelinsky}, \citenamefont {Zhang}, \citenamefont {Yu}, \citenamefont {Chen},\
  and\ \citenamefont {Samaras}}]{zelinsky2005role}%
  \BibitemOpen
  \bibfield  {author} {\bibinfo {author} {\bibfnamefont {G.}~\bibnamefont
  {Zelinsky}}, \bibinfo {author} {\bibfnamefont {W.}~\bibnamefont {Zhang}},
  \bibinfo {author} {\bibfnamefont {B.}~\bibnamefont {Yu}}, \bibinfo {author}
  {\bibfnamefont {X.}~\bibnamefont {Chen}}, \ and\ \bibinfo {author}
  {\bibfnamefont {D.}~\bibnamefont {Samaras}},\ }in\ \href@noop {} {\emph
  {\bibinfo {booktitle} {Advances in neural information processing systems}}}\
  (\bibinfo {year} {2005})\ pp.\ \bibinfo {pages} {1569--1576}\BibitemShut
  {NoStop}%
\bibitem [{\citenamefont {Rhodes}\ \emph {et~al.}(2014)\citenamefont {Rhodes},
  \citenamefont {Kello},\ and\ \citenamefont {Kerster}}]{rhodes2014intrinsic}%
  \BibitemOpen
  \bibfield  {author} {\bibinfo {author} {\bibfnamefont {T.}~\bibnamefont
  {Rhodes}}, \bibinfo {author} {\bibfnamefont {C.~T.}\ \bibnamefont {Kello}}, \
  and\ \bibinfo {author} {\bibfnamefont {B.}~\bibnamefont {Kerster}},\
  }\href@noop {} {\bibfield  {journal} {\bibinfo  {journal} {Visual Cognition}\
  }\textbf {\bibinfo {volume} {22}},\ \bibinfo {pages} {809} (\bibinfo {year}
  {2014})}\BibitemShut {NoStop}%
\bibitem [{\citenamefont {Kello}\ \emph {et~al.}(2007)\citenamefont {Kello},
  \citenamefont {Beltz}, \citenamefont {Holden},\ and\ \citenamefont
  {Van~Orden}}]{kello2007emergent}%
  \BibitemOpen
  \bibfield  {author} {\bibinfo {author} {\bibfnamefont {C.~T.}\ \bibnamefont
  {Kello}}, \bibinfo {author} {\bibfnamefont {B.~C.}\ \bibnamefont {Beltz}},
  \bibinfo {author} {\bibfnamefont {J.~G.}\ \bibnamefont {Holden}}, \ and\
  \bibinfo {author} {\bibfnamefont {G.~C.}\ \bibnamefont {Van~Orden}},\
  }\href@noop {} {\bibfield  {journal} {\bibinfo  {journal} {Journal of
  Experimental Psychology: General}\ }\textbf {\bibinfo {volume} {136}},\
  \bibinfo {pages} {551} (\bibinfo {year} {2007})}\BibitemShut {NoStop}%
\bibitem [{\citenamefont {Kello}\ \emph {et~al.}(2008)\citenamefont {Kello},
  \citenamefont {Anderson}, \citenamefont {Holden},\ and\ \citenamefont
  {Van~Orden}}]{kello2008pervasiveness}%
  \BibitemOpen
  \bibfield  {author} {\bibinfo {author} {\bibfnamefont {C.~T.}\ \bibnamefont
  {Kello}}, \bibinfo {author} {\bibfnamefont {G.~G.}\ \bibnamefont {Anderson}},
  \bibinfo {author} {\bibfnamefont {J.~G.}\ \bibnamefont {Holden}}, \ and\
  \bibinfo {author} {\bibfnamefont {G.~C.}\ \bibnamefont {Van~Orden}},\
  }\href@noop {} {\bibfield  {journal} {\bibinfo  {journal} {Cognitive
  Science}\ }\textbf {\bibinfo {volume} {32}},\ \bibinfo {pages} {1217}
  (\bibinfo {year} {2008})}\BibitemShut {NoStop}%
\bibitem [{\citenamefont {Kello}(2013)}]{kello2013critical}%
  \BibitemOpen
  \bibfield  {author} {\bibinfo {author} {\bibfnamefont {C.~T.}\ \bibnamefont
  {Kello}},\ }\href@noop {} {\bibfield  {journal} {\bibinfo  {journal}
  {Psychological review}\ }\textbf {\bibinfo {volume} {120}},\ \bibinfo {pages}
  {230} (\bibinfo {year} {2013})}\BibitemShut {NoStop}%
\bibitem [{\citenamefont {Sornette}(2006)}]{sornette2006critical}%
  \BibitemOpen
  \bibfield  {author} {\bibinfo {author} {\bibfnamefont {D.}~\bibnamefont
  {Sornette}},\ }\href@noop {} {\emph {\bibinfo {title} {Critical phenomena in
  natural sciences: chaos, fractals, selforganization and disorder: concepts
  and tools}}}\ (\bibinfo  {publisher} {Springer Science \& Business Media},\
  \bibinfo {year} {2006})\BibitemShut {NoStop}%
\bibitem [{\citenamefont {Feder}(2013)}]{feder2013fractals}%
  \BibitemOpen
  \bibfield  {author} {\bibinfo {author} {\bibfnamefont {J.}~\bibnamefont
  {Feder}},\ }\href@noop {} {\emph {\bibinfo {title} {Fractals}}}\ (\bibinfo
  {publisher} {Springer Science \& Business Media},\ \bibinfo {year}
  {2013})\BibitemShut {NoStop}%
\bibitem [{\citenamefont {Coey}\ \emph {et~al.}(2012)\citenamefont {Coey},
  \citenamefont {Wallot}, \citenamefont {Richardson},\ and\ \citenamefont
  {Van~Orden}}]{vanorden}%
  \BibitemOpen
  \bibfield  {author} {\bibinfo {author} {\bibfnamefont {C.~A.}\ \bibnamefont
  {Coey}}, \bibinfo {author} {\bibfnamefont {S.}~\bibnamefont {Wallot}},
  \bibinfo {author} {\bibfnamefont {M.~J.}\ \bibnamefont {Richardson}}, \ and\
  \bibinfo {author} {\bibfnamefont {G.}~\bibnamefont {Van~Orden}},\ }\href@noop
  {} {\bibfield  {journal} {\bibinfo  {journal} {Journal of Eye Movement
  Research}\ }\textbf {\bibinfo {volume} {5}},\ \bibinfo {pages} {1} (\bibinfo
  {year} {2012})}\BibitemShut {NoStop}%
\bibitem [{\citenamefont {Kantelhardt}\ \emph {et~al.}(2002)\citenamefont
  {Kantelhardt}, \citenamefont {Zschiegner}, \citenamefont {Koscielny-Bunde},
  \citenamefont {Havlin}, \citenamefont {Bunde},\ and\ \citenamefont
  {Stanley}}]{kantelhardt2002multifractal}%
  \BibitemOpen
  \bibfield  {author} {\bibinfo {author} {\bibfnamefont {J.~W.}\ \bibnamefont
  {Kantelhardt}}, \bibinfo {author} {\bibfnamefont {S.~A.}\ \bibnamefont
  {Zschiegner}}, \bibinfo {author} {\bibfnamefont {E.}~\bibnamefont
  {Koscielny-Bunde}}, \bibinfo {author} {\bibfnamefont {S.}~\bibnamefont
  {Havlin}}, \bibinfo {author} {\bibfnamefont {A.}~\bibnamefont {Bunde}}, \
  and\ \bibinfo {author} {\bibfnamefont {H.~E.}\ \bibnamefont {Stanley}},\
  }\href@noop {} {\bibfield  {journal} {\bibinfo  {journal} {Physica A:
  Statistical Mechanics and its Applications}\ }\textbf {\bibinfo {volume}
  {316}},\ \bibinfo {pages} {87} (\bibinfo {year} {2002})}\BibitemShut
  {NoStop}%
\bibitem [{\citenamefont {Kantelhardt}(2012)}]{kantelhardt2012fractal}%
  \BibitemOpen
  \bibfield  {author} {\bibinfo {author} {\bibfnamefont {J.~W.}\ \bibnamefont
  {Kantelhardt}},\ }in\ \href@noop {} {\emph {\bibinfo {booktitle} {Mathematics
  of complexity and dynamical systems}}}\ (\bibinfo  {publisher} {Springer},\
  \bibinfo {year} {2012})\ pp.\ \bibinfo {pages} {463--487}\BibitemShut
  {NoStop}%
\bibitem [{\citenamefont {Movahed}\ \emph {et~al.}(2006)\citenamefont
  {Movahed}, \citenamefont {Jafari}, \citenamefont {Ghasemi}, \citenamefont
  {Rahvar},\ and\ \citenamefont {Tabar}}]{movahed2006multifractal}%
  \BibitemOpen
  \bibfield  {author} {\bibinfo {author} {\bibfnamefont {M.~S.}\ \bibnamefont
  {Movahed}}, \bibinfo {author} {\bibfnamefont {G.}~\bibnamefont {Jafari}},
  \bibinfo {author} {\bibfnamefont {F.}~\bibnamefont {Ghasemi}}, \bibinfo
  {author} {\bibfnamefont {S.}~\bibnamefont {Rahvar}}, \ and\ \bibinfo {author}
  {\bibfnamefont {M.~R.~R.}\ \bibnamefont {Tabar}},\ }\href@noop {} {\bibfield
  {journal} {\bibinfo  {journal} {Journal of Statistical Mechanics: Theory and
  Experiment}\ }\textbf {\bibinfo {volume} {2006}},\ \bibinfo {pages} {P02003}
  (\bibinfo {year} {2006})}\BibitemShut {NoStop}%
\bibitem [{\citenamefont {Shang}\ \emph {et~al.}(2008)\citenamefont {Shang},
  \citenamefont {Lu},\ and\ \citenamefont {Kamae}}]{shang2008detecting}%
  \BibitemOpen
  \bibfield  {author} {\bibinfo {author} {\bibfnamefont {P.}~\bibnamefont
  {Shang}}, \bibinfo {author} {\bibfnamefont {Y.}~\bibnamefont {Lu}}, \ and\
  \bibinfo {author} {\bibfnamefont {S.}~\bibnamefont {Kamae}},\ }\href@noop {}
  {\bibfield  {journal} {\bibinfo  {journal} {Chaos, Solitons \& Fractals}\
  }\textbf {\bibinfo {volume} {36}},\ \bibinfo {pages} {82} (\bibinfo {year}
  {2008})}\BibitemShut {NoStop}%
\bibitem [{\citenamefont {Lim}\ \emph {et~al.}(2007)\citenamefont {Lim},
  \citenamefont {Kim}, \citenamefont {Lee}, \citenamefont {Kim},\ and\
  \citenamefont {Lee}}]{lim2007multifractal}%
  \BibitemOpen
  \bibfield  {author} {\bibinfo {author} {\bibfnamefont {G.}~\bibnamefont
  {Lim}}, \bibinfo {author} {\bibfnamefont {S.}~\bibnamefont {Kim}}, \bibinfo
  {author} {\bibfnamefont {H.}~\bibnamefont {Lee}}, \bibinfo {author}
  {\bibfnamefont {K.}~\bibnamefont {Kim}}, \ and\ \bibinfo {author}
  {\bibfnamefont {D.-I.}\ \bibnamefont {Lee}},\ }\href@noop {} {\bibfield
  {journal} {\bibinfo  {journal} {Physica A: Statistical Mechanics and its
  Applications}\ }\textbf {\bibinfo {volume} {386}},\ \bibinfo {pages} {259}
  (\bibinfo {year} {2007})}\BibitemShut {NoStop}%
\bibitem [{\citenamefont {Yuan}\ \emph {et~al.}(2009)\citenamefont {Yuan},
  \citenamefont {Zhuang},\ and\ \citenamefont {Jin}}]{yuan2009measuring}%
  \BibitemOpen
  \bibfield  {author} {\bibinfo {author} {\bibfnamefont {Y.}~\bibnamefont
  {Yuan}}, \bibinfo {author} {\bibfnamefont {X.-t.}\ \bibnamefont {Zhuang}}, \
  and\ \bibinfo {author} {\bibfnamefont {X.}~\bibnamefont {Jin}},\ }\href@noop
  {} {\bibfield  {journal} {\bibinfo  {journal} {Physica A: Statistical
  Mechanics and its Applications}\ }\textbf {\bibinfo {volume} {388}},\
  \bibinfo {pages} {2189} (\bibinfo {year} {2009})}\BibitemShut {NoStop}%
\bibitem [{\citenamefont {Wang}\ \emph {et~al.}(2009)\citenamefont {Wang},
  \citenamefont {Liu},\ and\ \citenamefont {Gu}}]{wang2009analysis}%
  \BibitemOpen
  \bibfield  {author} {\bibinfo {author} {\bibfnamefont {Y.}~\bibnamefont
  {Wang}}, \bibinfo {author} {\bibfnamefont {L.}~\bibnamefont {Liu}}, \ and\
  \bibinfo {author} {\bibfnamefont {R.}~\bibnamefont {Gu}},\ }\href@noop {}
  {\bibfield  {journal} {\bibinfo  {journal} {International Review of Financial
  Analysis}\ }\textbf {\bibinfo {volume} {18}},\ \bibinfo {pages} {271}
  (\bibinfo {year} {2009})}\BibitemShut {NoStop}%
\bibitem [{\citenamefont {Galaska}\ \emph {et~al.}(2008)\citenamefont
  {Galaska}, \citenamefont {Makowiec}, \citenamefont {Dudkowska}, \citenamefont
  {Koprowski}, \citenamefont {Chlebus}, \citenamefont {Wdowczyk-Szulc},\ and\
  \citenamefont {Rynkiewicz}}]{galaska2008comparison}%
  \BibitemOpen
  \bibfield  {author} {\bibinfo {author} {\bibfnamefont {R.}~\bibnamefont
  {Galaska}}, \bibinfo {author} {\bibfnamefont {D.}~\bibnamefont {Makowiec}},
  \bibinfo {author} {\bibfnamefont {A.}~\bibnamefont {Dudkowska}}, \bibinfo
  {author} {\bibfnamefont {A.}~\bibnamefont {Koprowski}}, \bibinfo {author}
  {\bibfnamefont {K.}~\bibnamefont {Chlebus}}, \bibinfo {author} {\bibfnamefont
  {J.}~\bibnamefont {Wdowczyk-Szulc}}, \ and\ \bibinfo {author} {\bibfnamefont
  {A.}~\bibnamefont {Rynkiewicz}},\ }\href@noop {} {\bibfield  {journal}
  {\bibinfo  {journal} {Annals of Noninvasive Electrocardiology}\ }\textbf
  {\bibinfo {volume} {13}},\ \bibinfo {pages} {155} (\bibinfo {year}
  {2008})}\BibitemShut {NoStop}%
\bibitem [{\citenamefont {Zorick}\ and\ \citenamefont
  {Mandelkern}(2013)}]{zorick2013multifractal}%
  \BibitemOpen
  \bibfield  {author} {\bibinfo {author} {\bibfnamefont {T.}~\bibnamefont
  {Zorick}}\ and\ \bibinfo {author} {\bibfnamefont {M.~A.}\ \bibnamefont
  {Mandelkern}},\ }\href@noop {} {\bibfield  {journal} {\bibinfo  {journal}
  {PloS one}\ }\textbf {\bibinfo {volume} {8}},\ \bibinfo {pages} {e68360}
  (\bibinfo {year} {2013})}\BibitemShut {NoStop}%
\bibitem [{\citenamefont {Zhang}\ \emph {et~al.}(2008)\citenamefont {Zhang},
  \citenamefont {Xu}, \citenamefont {Chen},\ and\ \citenamefont
  {Yu}}]{zhang2008multifractal}%
  \BibitemOpen
  \bibfield  {author} {\bibinfo {author} {\bibfnamefont {Q.}~\bibnamefont
  {Zhang}}, \bibinfo {author} {\bibfnamefont {C.-Y.}\ \bibnamefont {Xu}},
  \bibinfo {author} {\bibfnamefont {Y.~D.}\ \bibnamefont {Chen}}, \ and\
  \bibinfo {author} {\bibfnamefont {Z.}~\bibnamefont {Yu}},\ }\href@noop {}
  {\bibfield  {journal} {\bibinfo  {journal} {Hydrological Processes}\ }\textbf
  {\bibinfo {volume} {22}},\ \bibinfo {pages} {4997} (\bibinfo {year}
  {2008})}\BibitemShut {NoStop}%
\bibitem [{\citenamefont {Telesca}\ and\ \citenamefont
  {Lovallo}(2011)}]{telesca2011analysis}%
  \BibitemOpen
  \bibfield  {author} {\bibinfo {author} {\bibfnamefont {L.}~\bibnamefont
  {Telesca}}\ and\ \bibinfo {author} {\bibfnamefont {M.}~\bibnamefont
  {Lovallo}},\ }\href@noop {} {\bibfield  {journal} {\bibinfo  {journal}
  {Journal of Statistical Mechanics: Theory and Experiment}\ }\textbf {\bibinfo
  {volume} {2011}},\ \bibinfo {pages} {P07001} (\bibinfo {year}
  {2011})}\BibitemShut {NoStop}%
\bibitem [{\citenamefont {Findlay}\ and\ \citenamefont
  {Walker}(1999)}]{Findlay}%
  \BibitemOpen
  \bibfield  {author} {\bibinfo {author} {\bibfnamefont {J.~M.}\ \bibnamefont
  {Findlay}}\ and\ \bibinfo {author} {\bibfnamefont {R.}~\bibnamefont
  {Walker}},\ }\href@noop {} {\bibfield  {journal} {\bibinfo  {journal}
  {Behavioral and Brain Sciences}\ }\textbf {\bibinfo {volume} {22}},\ \bibinfo
  {pages} {661} (\bibinfo {year} {1999})}\BibitemShut {NoStop}%
\bibitem [{\citenamefont {Itti}\ \emph {et~al.}(1998)\citenamefont {Itti},
  \citenamefont {Koch},\ and\ \citenamefont {Niebur}}]{Itti}%
  \BibitemOpen
  \bibfield  {author} {\bibinfo {author} {\bibfnamefont {L.}~\bibnamefont
  {Itti}}, \bibinfo {author} {\bibfnamefont {C.}~\bibnamefont {Koch}}, \ and\
  \bibinfo {author} {\bibfnamefont {E.}~\bibnamefont {Niebur}},\ }\href@noop {}
  {\bibfield  {journal} {\bibinfo  {journal} {IEEE Transactions on Pattern
  Analysis and Machine Intelligence}\ }\textbf {\bibinfo {volume} {20}},\
  \bibinfo {pages} {1254} (\bibinfo {year} {1998})}\BibitemShut {NoStop}%
\bibitem [{\citenamefont {Lamy}\ \emph {et~al.}(2004)\citenamefont {Lamy},
  \citenamefont {Leber},\ and\ \citenamefont {Egeth}}]{Lamy}%
  \BibitemOpen
  \bibfield  {author} {\bibinfo {author} {\bibfnamefont {D.}~\bibnamefont
  {Lamy}}, \bibinfo {author} {\bibfnamefont {A.}~\bibnamefont {Leber}}, \ and\
  \bibinfo {author} {\bibfnamefont {H.~E.}\ \bibnamefont {Egeth}},\ }\href@noop
  {} {\bibfield  {journal} {\bibinfo  {journal} {Journal of Experimental
  Psychology: Human Perception and Performance}\ }\textbf {\bibinfo {volume}
  {30}},\ \bibinfo {pages} {1254} (\bibinfo {year} {2004})}\BibitemShut
  {NoStop}%
\bibitem [{\citenamefont {Land}\ and\ \citenamefont {Hayhoe}(2001)}]{Land}%
  \BibitemOpen
  \bibfield  {author} {\bibinfo {author} {\bibfnamefont {M.}~\bibnamefont
  {Land}}\ and\ \bibinfo {author} {\bibfnamefont {M.}~\bibnamefont {Hayhoe}},\
  }\href@noop {} {\bibfield  {journal} {\bibinfo  {journal} {Vision Research}\
  }\textbf {\bibinfo {volume} {41}},\ \bibinfo {pages} {3559} (\bibinfo {year}
  {2001})}\BibitemShut {NoStop}%
\bibitem [{\citenamefont {Underwood}\ and\ \citenamefont
  {Foulsham}(2006)}]{Underwood}%
  \BibitemOpen
  \bibfield  {author} {\bibinfo {author} {\bibfnamefont {G.}~\bibnamefont
  {Underwood}}\ and\ \bibinfo {author} {\bibfnamefont {T.}~\bibnamefont
  {Foulsham}},\ }\href@noop {} {\bibfield  {journal} {\bibinfo  {journal} {The
  Quarterly Journal of Experimental Psychology}\ }\textbf {\bibinfo {volume}
  {59}},\ \bibinfo {pages} {1931} (\bibinfo {year} {2006})}\BibitemShut
  {NoStop}%
\bibitem [{\citenamefont {Rothkopf}\ \emph {et~al.}(2007)\citenamefont
  {Rothkopf}, \citenamefont {Ballard},\ and\ \citenamefont {Hayhoe}}]{Ballard}%
  \BibitemOpen
  \bibfield  {author} {\bibinfo {author} {\bibfnamefont {C.}~\bibnamefont
  {Rothkopf}}, \bibinfo {author} {\bibfnamefont {D.}~\bibnamefont {Ballard}}, \
  and\ \bibinfo {author} {\bibfnamefont {M.}~\bibnamefont {Hayhoe}},\
  }\href@noop {} {\bibfield  {journal} {\bibinfo  {journal} {Journal of
  Vision}\ }\textbf {\bibinfo {volume} {7}},\ \bibinfo {pages} {1} (\bibinfo
  {year} {2007})}\BibitemShut {NoStop}%
\bibitem [{\citenamefont {Hale}\ \emph {et~al.}(2011)\citenamefont {Hale},
  \citenamefont {Hawkins}, \citenamefont {Sheeley}, \citenamefont {Reynolds},
  \citenamefont {Jenkins}, \citenamefont {Schmitt},\ and\ \citenamefont
  {Martin}}]{PITS:PITS20543}%
  \BibitemOpen
  \bibfield  {author} {\bibinfo {author} {\bibfnamefont {A.~D.}\ \bibnamefont
  {Hale}}, \bibinfo {author} {\bibfnamefont {R.~O.}\ \bibnamefont {Hawkins}},
  \bibinfo {author} {\bibfnamefont {W.}~\bibnamefont {Sheeley}}, \bibinfo
  {author} {\bibfnamefont {J.~R.}\ \bibnamefont {Reynolds}}, \bibinfo {author}
  {\bibfnamefont {S.}~\bibnamefont {Jenkins}}, \bibinfo {author} {\bibfnamefont
  {A.~J.}\ \bibnamefont {Schmitt}}, \ and\ \bibinfo {author} {\bibfnamefont
  {D.~A.}\ \bibnamefont {Martin}},\ }\href {\doibase 10.1002/pits.20543}
  {\bibfield  {journal} {\bibinfo  {journal} {Psychology in the Schools}\
  }\textbf {\bibinfo {volume} {48}},\ \bibinfo {pages} {4} (\bibinfo {year}
  {2011})}\BibitemShut {NoStop}%
\end{thebibliography}%

\end{document}